# Forecasting exports in selected OECD countries and Iran using MLP Artificial Neural Network

**Authors Details:**

**-** Soheila Khajoui
- Saeid Dehyadegari
- Sayyed Abdolmajid Jalaee

**Affiliations**
- Department of Management, Faculty of Management & Economics, Shahid Bahonar University of Kerman, Kerman, Iran
- Department of Management, Faculty of Management & Economics, Shahid Bahonar University of Kerman, Kerman
- Department of Economics, Faculty of Management & Economics, Shahid Bahonar University of Kerman, Kerman, Iran

**Author's email address**

- s.khajoui@aem.uk.ac.ir

- dehyadegari@uk.ac.ir

- jalaee@uk.ac.ir

**Corresponding author:**

Saeid Dehyadegari:

- Department of Management, Faculty of Management & Economics, Shahid Bahonar University of Kerman, Kerman, Iran
- dehyadegari@uk.ac.ir

**Declaration of Conflicting Interests**
The authors declared no potential conflicts of interest with respect to the research, authorship and/or publication of this article.

**Funding**
The authors received no financial support for the research, authorship and/or publication of this article.

**Abstract**

The present study aimed to forecast the exports of a select group of Organization for Economic Co-operation and Development (OECD) countries and Iran using the neural networks. The data concerning the exports of the above countries from 1970 to 2019 were collected. The collected data were implemented to forecast the exports of the investigated countries for 2021 to 2025. The analysis was performed using the Multi-Layer-Perceptron (MLP) neural network in Python. Out of the total number, 75 percent were used as training data, and 25 percent were used as the test data. The findings of the study were evaluated with 99% accuracy, which indicated the reliability of the output of the network. The Results show that Covid-19 has affected exports over time. However, long-term export contracts are less affected by tensions and crises, due to the effect of exports on economic growth, per capita income and it is better for economic policies of countries to use long-term export contracts.

**JEL Classification:** F17, B26, C45

**Keywords**: Artificial intelligence, Export Forecasting, International Trade, Multi-layer perceptron neural network, modeling

## Introduction

Due to the development of artificial intelligence and neural network in international trade as well as its effective use in forecasting, artificial neural networks have been used to predict businesses, including exports in recent years. As a result, MLP artificial neural network algorithms used in this study specifically for export forecasts, which showed higher accuracy for export prediction than other neural network methods, including Genetic programming (GP). In predicting the nonlinear export system, the rationality and functionality of the model's construction has the benefits that can create a nonlinear prediction system for export and provide a wider space for development.

Trading between countries has a long history and is considered as one of the characteristics of human societies (Hulme, 2021). Since old times, countries have exchanged internationally to meet their needs and obtain the required commodities and services. Moreover, eliminating the obstacles against international trade and the increased trans-border exchanges like exports, direct the economic resources of different nations toward activities with higher productivity and accelerates their development and advancement (Salehi, 2016).

Technology advancement has infiltrated the field of international trade and has resulted in its enhancement and development, so that millions of people around the world use the Internet to do anything from doing research to shopping online. As export increases the survival chances of enterprises (particularly in developing countries), and enhances their shares of the opportunities provided by international markets, it is considered a very important topic for them (Bagheri Kahyesh & Bakhshandeh, 2021). Indeed, export is among the most fundamental and important branches of the economy of each country (Tafaghodi et al., 2020). International trade (including export) is typically performed by national governments (Garbellini, 2021). Predicting the status of exports assists governments and enterprises in better policy-making and decision-making, assessing the rate of productivity, and getting prepared for international trade in the future (Dave et al., 2021). Thus, predicting exports plays a significant role in organizations as a way of investigating the relationship between internal decisions and uncontrollable external factors.

The traditional methods of time-series prediction work according to the existing trends between data and are unable to predict macro-economic changes and detect nonlinear and advanced patterns. Nevertheless, the neural networks as MLP perform conveniently in terms of predicting nonlinear and advanced patterns and bulky data and have attracted significant attention for these properties (Sohrabpour et al., 2021). MLP Neural Network is a remarkable non-linear multi-factor evaluation method and one of the most commonly used Ann models. MLP-Ann is suitable for solving practical issues because of its complex structure and is used as a comprehensive method for predicting international trade. (Ma et al., 2021). Therefore, the MLP neural network is used as a comprehensive method for export forecasting.

As export forms a significant portion of the international trade is an important factor for many economic enterprises, its predictions need special attention. Using such predicts can reduce the time, costs, and challenges that an enterprise will face in terms of trade – particularly exports. Moreover, obstacles to the collection of data including the issues of availability, the lack of high-quality data, the significant size of the data, and the requirement of the development of various models or methods according to the unique characteristics of countries constrain approaches used for prediction. Such factors turn the export sales into a distinct and challenging area that cannot be determined merely by traditional methods like the time series to maintain empirical accuracy and dependability. Thus, the present study implemented the neural network method due to its high accuracy and dependability in addition to its potential in detecting nonlinear relationships. Though several studies were conducted on the prediction of exports, the present study could contribute to the existing literature as it implemented the Perceptron neural network (a method with high capabilities and accuracy in the prediction of time series) to predict the rates of export and investigate export trends in Iran and 9 selected OECD countries. Moreover, the selection of the OECD countries and Iran added to the significance and innovation of the present study as the samples included developed, developing, and underdeveloped countries.

**Literature review**

Predicting sales – particularly in exports – plays a significant role in linking internal decision-making and uncontrollable external factors and can influence organizations. There are various methods for predicting sales. Typically, the prediction of the demand and sales was performed using traditional time-series methods like ARIMA and the Holt-Winters method. However, newer artificial intelligence (AI-based) methods have gained prominence due to their capabilities in increasing the efficiency of prediction and the modeling of nonlinear patterns. For instance, some studies have implemented the GP method to predict the rate of exports. Combining AI with trade methods seems necessary due to the emergence of large and advanced data (Sohrabpour et al., 2021).

AI has enables organizations to know their customers better and effectively target them using customized digital messages; thus, many suppliers of financial services implement AI to enhance trade methods (Mogaji, Soetan & Kieu, 2020). Predictions regarding the OECD trade can be improved using standard time-series methods rather than the simpler ones (Keck, Raubold & Truppia, 2009).

Export and import are two macro-economic variables that have gained significant importance due to universal relationships and the optimal allocation of resources (Shahabadi, Amiri & Sarigol, 2016).

**Model**

International trade is an indispensable part of the cost-reduction strategies of large original equipment manufacturer (OEMs) that are placed at the top of global value chains (Garbellini, 2021). International trade includes the trans-border flow of capital and facilitates the global exchange of commodities and services by reducing the financial obstacles to business transactions and enhancing commercial investments (Lai, Wang & Xu, 2021). Companies can enter international markets, increase their sales, and obtain the required information and knowledge via the channel of export. Moreover, exports increase economic growth and influence employment and the balance of wages (Bagheri Kahyesh & Bakhshandeh, 2021).

Modern technological developments have prompted industries to step on the path of technology and have also resulted in the evolution and enhancement of marketing. Indeed, all systems that work according to AI facilitate decision-making processes and replace traditional methods (Dumitriu & Popescu, 2020). Traditionally, salespeople needed human resources in areas like sales, marketing research, the identification of customers and competitors, and advertisement, but the emergence of AI and neural networks for such affairs minimized the need for humans (Paschen, Wilson & Ferreira, 2020).

Most AI applications in the field of business are based on using artificial neural networks to perform complex tasks that were considered unsolvable less than a decade ago. Marketing predictions enable marketers to obtain a view of the future marketing activities and understand their influence on enhancing procedures, recruiting new customers, and achieving optimal customers (De Bruyn et al., 2020). Neural networks are somehow similar to the linear and nonlinear regression models and can be used instead of the typical statistical approaches as they are more reliable than other methods. The most prevalent artificial neural network is the Perceptron network, which consists of three layers including the input layer, the hidden layer, and the output layer (Tohidi et al., 2015).

The demand for exporting a certain commodity is influenced by factors like its export price, the global export price, its average price in international markets, and the income level of the importing countries. Thus, the export demand function can be defined in the following logarithmic form:

$$ln\, X_t^d = a_0 + a_1 \ln PX_t + a_2 \ln PXW_t + a_3 \ln YW_t + a_4\, \ln W_t + U_{1t} \tag{1}$$

Where $X^d$ is the global demand for export, PX is the export price index of the commodity, PXW is the global price of export, YW is the average weighted and actual income of the importing countries, W is the rate of the production of that commodity in other countries, and U1 is the disturbance term.

In general, the supply of export depends on factors like the export price of the commodity and its domestic price and production. Thus, the export supply function can be written in the following logarithmic form:

$$ln\, X_t^s = \beta_0 + \beta_1 \ln(PX/P)_t + \beta_2 \ln Y_t + U_{4t} \tag{2}$$

Where $X_t^s$ is the export supply rate, PX is the export price of the commodity, P is the domestic price of the good, $Y_t$ is the rate of its production inside the country, and $U_{4t}$ is the disturbance term (Goldstein & Khan, 1978).

### Method

The present study was an applied descriptive one that was conducted using time-series data and the MLP neural network. Data collection was based on time-series data, and this was carried out by checking various articles, resources, and credible international websites like (among others) the World Bank, World Trade Organization (WTO), and IFM.

For this purpose, first, the export data of the U.S., Canada, Germany, France, Japan, Turkey, South Korea, Portugal, Greece, and Iran related to a 50-year period (1970-2019) were collected from the world bank database and were transformed into seasonal data using Diz's formula (resulting in 200 pieces of data for each country). That was because when more data were available, the performance of the neural network got more favorable. Predicting the rate of exports in Iran and OECD members including the U.S., Canada, Germany, France, Japan, Turkey, South Korea, Portugal, and Greece was conducted for 5 years (20 seasons) during the period from 2021 to 2025. The selection of a sample that included developed, developing, and underdeveloped countries added to the importance and innovativeness of the study. The method of analysis was based on neural network algorithms, and the implemented MLP network is explained below in terms of its structure and performance.

### Multi-layer Perceptron neural network

Using neural networks has multiple advantages over the traditional methods of prediction and makes it possible to analyze a large amount of data in very little time (Desai and Shah 2021). The most prevalent form of neural networks is the multi-layer Perceptron network, which consists of an input layer, a hidden layer, and an output layer (Figure 1).

The input layer sends messages to the nodes in the hidden layers. The latter weights and collects the received messages. Each node in the hidden layer is a function that collects the incoming weights and sends them to the output layer (Ma et al., 2021). (Figure 2).

Assume that G is the number of hidden layers in the MLP network, and the neurons of each layer are $n_0, n_1, \ldots, n_{G-1}$, respectively. In this way, the input and output of the first hidden layer are calculated according to the following equations:

$$i_n = \sum_{j=1}^{n_0} w_{jn} f_0(x_j) + b_n \qquad (3)$$

$$o_n = f_1(i_n) = \left(1 + e^{-i_n}\right)^{-1} \qquad (4)$$

Where $i_n$ and $o_n$ are the input and output of the first hidden layer, respectively, $b_n$ is the threshold or the bias of the *n*th neuron of the first hidden layer. Moreover, $w_{jn}$ is the weighted value between the *j*th incoming neuron and the nth neuron in the first layer, and $f_k(0)$ is the activation function of the kth layer that determines the relationships between the input and output of the neurons and the network. Thus, the input and output values of the output layers are determined in the following manner:

$$i_1 = \sum_{j=1}^{n_G} w_{j1} o_j + b_1 \qquad (5)$$

$$o_1 = i_1 \qquad (6)$$

Where, $i_1$, $o_1$, and $b_1$ are the input, output, and the bias of neurons in the output layer, respectively, and $w_{j1}$ is the weight between the *j*th neuron in the hidden layer and the neuron of the output layer.

The neural network consists of two main processes including training and testing. Neural networks are trained to perform specified tasks by the modification of weights. The weights are constantly modified by comparing the actual and predicted outputs until the error function, which indicates the difference between the actual and predicted outputs, is minimized. Moreover, the process is determined based by entering the input for each epoch based on average values (Batch-Learning), and the value for each epoch was determined at 200 in the present study. There are multiple methods based on the gradient descent algorithm and the Newton method to minimize the error function. However, the error backpropagation method is the most prevalent one in this regard. The error backpropagation algorithm acts according to the estimation of the gradient descent. In the algorithm, the weights of the network move in a reverse direction relative to the gradient of the efficiency function. Indeed, after the propagation of the minimum, the error performance function implemented the gradient descent method, which is considered a classical optimization approach. The index of efficiency in this algorithm is the mean squared error (MSE). After the process of training is over, the process of testing starts on the relevant data, and the MLP neural network is trained using a relu activation function and the adam optimizer (Martínez-Comesaña et al., 2021).

### Results

The MLP neural network that was designed in the present study consisted of an input layer, a hidden layer, and an output layer. The network was implemented several times using various neurons and parameters Run on Python software, and the most desirable state was the one where the MSE and MAE indices were the least. The MSE and absolute error (MAE) results of the training and test data used to predict the exports of each investigated country based on the normalized data are provided in tables 1 and 2. 75 percent of the input data were considered the training data, while the remaining 25 percent were considered the test data. The number of EPOCH 200 was considered. The network also uses a Relu activaton function and ADAM optimizer to teach data. The libraries used in this software include Numpy, Tensorflow, Keras, Panda and Matplotlib. Finally, the results of the data analysis to predict the exports of the selected countries for 20 seasons (2021-25) were obtained with an accuracy of 99 percent which is provided in table 3. (q is season).

**Normalization of data**

Neural network algorithms work better or converge faster when different characteristics (variables) are smaller; Therefore, data normalization is done before training neural network models. In this process, conversions are made to the inputs and outputs of the network to remove the features from the inputs and convert the output to a more understandable way for the network. This will prevent computational problems.

There are different methods for normalization, one of which is the following methods, in which the data falls between 0 and 1.

$$x_{norm} = \frac{x - x_{min}}{x_{max} - x_{min}} \tag{7}$$

In this formula X indicates a variable value  (Kumar &  Gupta,2010)

**Training and Test data**

The training and test data is usually needed to build a neural network in line with the purpose of predicting. The training data is used to estimate network weights and, in fact, create neural network model and training, and the test data is adopted to evaluate the ability to generalize the model or in other words to evaluate the ability to predict the model. Sometimes a third data, called the evaluation data, is used to prevent more of a fit problem or to determine the stoppage of the training process. But the use of a single data for the purpose of testing and evaluation is especially common in small data set (Zhang et al., 1998).

The important issue is how the data is divided into these two sets. In this regard, issues such as the characteristics of the problem and the type and size of the available data should be considered. These two sets must have the characteristics of society. This is very important in predicting time series. The existing literature provides little guidance in choosing the test and training data. In general, the richer the data is, the more accurate the prediction results are probably) (Zarra nejad et al., 2008).

**Error measurement criteria**

Different criteria are used to evaluate the predicted accuracy by different methods. The criteria that are most commonly used are the mean square squares (MSE), the root of the error squares (RMSE), the percentage of absolute error (MAPE), and the absolute error (MAE).

MSE and RMSE criteria are the most applicable performance evaluation index of different prediction methods. This index has been used in various studies. The average MSE calculates the error rate per view.

$$MSE = \frac{1}{T}\sum_{t=1}^{T}(P-A)^2 \tag{8}$$

The root of the error squares (RMSE) is the result of the absorption of the mean square squares.

$$RMSE = \sqrt{\frac{1}{T}\sum_{t=1}^{T}(P-A)^2} \tag{9}$$

The MAPE criterion represents a percentage error and is the most popular criterion without unit.

$$MAPE = \frac{1}{T}\sum_{t=1}^{T}\left|\frac{P-A}{A}\right| * 100 \tag{10}$$

The MAE criterion indicates the average amount of error, whether positive or negative.

$$MAE = \frac{1}{T}\sum_{t=1}^{T}|P-A| \tag{11}$$

In the above equations P and A, respectively, indicate predicted values and values are real. The less prediction error criteria indicate the high precision of prediction (Saifol Hosseini, Mohammadi Nejad, Moghaddasi, 2015).

**k-fold cross-validation**

In this case, the data is classified into K under the fold or fold. One of these parts is separated and fitted with the rest of the model data. The model error is measured by the separated part. This is repeated to apply all parts or folds to measure errors. The average errors will be the model error estimation. In this study, the average accuracy is 99 %.

Based on the results obtained for the export amount of the selected OECD countries and Iran (Table 3), the predicted export trends for the selected OECD countries and Iran for a 20-season period (2021-2025) were as follows in Figure 3.

In the following, the charts related to the MSE and the predicted data including the training and test data and their regression analyses for Iran are specified. The results showed that the regression coefficient between the targeted values and the output of the MLP network for all data was 0.99. (Figure 4-6)

Figure 7 and 8 shows that the accuracy of the export data forecast was acceptable in the study because the closer the points to the regression line, the higher the prediction accuracy and the accuracy in the present study is 99 %.

*Y1=0.9913217732561934\*X+0.001464295272312886*                            *(*12)
*Y2=0.994180802252052\*X+0.002534063870274497*                             (13)

Y1, Y2 represents export based on test, train data, and X represents train predict.

### Conclusion

In the past few years, traditional forecasting methods, including ARIMA, have been used to predict time series, which have shown that they have poorly predicted non -linear and mass data. For example, Alam (2019) states that to predict the annual import and export of Saudi Arabia, the traditional ARIMA method is not suitable for time series with a large amount of data, and in today's era, artificial neural networks (ANN) are the most used to predict time series with a large amount of data. Also, Urotia, Abdul, Atinza (2019) in research to forecast the import and export of the Philippines concluded that neural network methods are the most suitable model for forecasting. They also found that the best model in forecasting the import and export values from 2018 to 2022 is the neural network model compared to the results of the ARIMA model. Because the actual values and the predicted values of imports and exports using ANN have a small error, it means that our actual data and the predicted data match well enough. In addition, Saifol Hosseini et al. (2014) in a study aimed at comparing the predictive power of neural network methods and ARIMA in predicting Iran's skin and leather exports showed that ARIMA method is the weakest method compared to neural network methods and neural network method. It greatly reduces the prediction error. Hence, Due to the special structure that exists in the export function as a whole, both in developed and underdeveloped countries (in general, the relationships that exist in the export function are not inherently linear), it will be better to estimate and estimate methods, especially for predicting Non-linear method should be used.

For this reason, in order to forecast exports in selected OECD member countries and Iran, the MLP neural network method has been used in this study due to its flexibility and ability to model non-linear relationships, and the results of this method have high reliability. Export time series data were collected from 1970 to 2019 in selected countries; And after applying it to the network, the prediction accuracy of 99% was obtained. This research sought to answer the question of how the export trend in the selected OECD countries and Iran will be during the period of time that is part of the post-corona period, because the natural crisis such as Covid-19 has caused export to be severely damaged. For this reason, in this research, export forecasting has been done. Export is the first sector that is damaged during international crises, whether financial or economic, and natural crises. This effect is shown during the period of time. The results of the predicted outputs show that the export has been damaged by the current of the Covid-19 crisis, and a major part of the export share, which reacts immediately due to the short-term trend, can be easily corrected and return to the stable state. For this reason, the economic policy of these countries should go towards long-term agreements that are less affected by tensions, and considering the effect that export can have on economic growth, per capita income and welfare level, long-term agreements are better.

**References**

Alam, T. (2019). Forecasting exports and imports through artificial neural network and autoregressive integrated moving average. *Decision Science Letters*, 8(3), 249-260.

Bagheri Kahyesh, A. & Bakhshandeh, G. (2021). Influencing the Effect of Export Market Orientation on Export Performance: Mediating Role of Marketing Effectiveness and Marketing Capabilities (Case Study: Export Companies in Ahvaz). *Journal of Marketing Management,* 16(52), 127-141. (in Persian) doi: https://sid.ir/paper/410508/fa

Dave, E., Leonardo, A., Jeanice, M. & Hanafiah, N. (2021). Forecasting Indonesia exports using a hybrid model ARIMA-LSTM. *Procedia Computer Science,* 179: 480-487. doi: 10.1016/J.PROCS.2021.01.031.

De Bruyn, A., Viswanathan, V., Beh, Y. S., Brock, J. K. U. & Von Wangenheim, F. (2020). Artificial intelligence and marketing: Pitfalls and opportunities. *Journal of Interactive Marketing,* 51 (1), 91-105. doi: 10.1016/J.INTMAR.2020.04.007.

Desai, M., & Shah, M. (2021). An anatomization on breast cancer detection and diagnosis employing multi-layer perceptron neural network (MLP) and Convolutional neural network (CNN). *Clinical Health,* 4, 1-11. doi: 10.1016/J.CEH.2020.11.002.

Dumitriu, D. & Popescu, M. A. M. (2020). Artificial intelligence solutions for digital marketing. *Procedia Manufacturing*, 46, 630-636. doi: 10.1016/J.PROMFG.2020.03.090.

Garbellini, N. (2021). International trade as a process of choice of technique. *Structural Change and Economic Dynamics,* 59, 42-50. doi: 10.1016/J.STRUECO.2021.08.003.

Goldstein, M. & Khan, M. S. (1978). The supply and demand for exports: a simultaneous approach. *The Review of Economics and Statistics,* 60, 275–286. doi: 10.2307/1924981 / jstor.org/stable/1924981

Hoffmann, L. F. S., Bizarria, F. C. P. & Bizarria, J. W. P. (2020). Detection of liner surface defects in solid rocket motors using multilayer perceptron neural networks. *Polymer Testing*, 88, 106559. doi: 10.1016/J.POLYMERTESTING.2020.106559.

Hoffmann, L. F. S., Bizarria, F. C. P., & Bizarria, J. W. P. (2020). Detection of liner surface defects in solid rocket motors using multilayer perceptron neural networks. Polymer Testing, 88, 106559.

Hulme, P. E. (2021). Unwelcome exchange: International trade as a direct and indirect driver of biological invasions worldwide. *One Earth*, 4(5), 666-679. doi: 10. 1016/ J. ONEEAR. 2021. 04. 015.

Kaefer, F., Heilman, C. M. & Ramenofsky, S. D. (2005). A neural network application to consumer classification to improve the timing of direct marketing activities. *Computers & Operations Research*, 32 (10), 2595-2615. doi: 10. 1016/ J.COR. 2004.06.021.

Keck, A., Raubold, A. & Truppia, A. (2010). Forecasting international trade: A time series approach. *OECD Journal: Journal of Business Cycle Measurement and Analysis*, 2009(2), 157-176. doi: 10.1787/jbcma-2009-5ks9v44bdj32.

Kumar, G., & Gupta, S. (2010). Forecasting Exports of Industrial Goods from Punjab-An Application of Univariate ARIMA Model. Annals of the University of Petrosani, Economics, 10(4), 169-180.

Lai, K., Wang, T. & Xu, D. (2021). Capital controls and international trade: An industry financial vulnerability perspective. *Journal of International Money and Finance,* 116, 102399. doi: 10.1016/J.JIMONFIN.2021.102399.

Ma, Z., Li, X., Chen, Y., Tang, X., Gao, Y., Wang, H. & Liu, R. (2021). Comprehensive evaluation of the combined extracts of Epimedii Folium and Ligustri Lucidi Fructus for PMOP in ovariectomized rats based on MLP-ANN methods. *Journal of Ethnopharmacology,* 268, 113563.

Martínez-Comesaña, M., Ogando-Martínez, A., Troncoso-Pastoriza, F., López-Gómez, J., Febrero-Garrido, L. & Granada-Álvarez, E. (2021). Use of optimised MLP neural networks for spatiotemporal estimation of indoor environmental conditions of existing buildings. *Building and Environment*, 205, 108243. doi: 10. 1016/ J. BUILDENV. 2021. 108243.

Mogaji, E., Soetan, T. O. & Kieu, T. A. (2020). The implications of artificial intelligence on the digital marketing of financial services to vulnerable customers. *Australasian Marketing Journal*, 29, 235–242. doi: 10.1016/j.ausmj.2020.05.003.

Mustak, M., Salminen, J., Plé, L. & Wirtz, J. (2021). Artificial intelligence in marketing: Topic modeling, scientometric analysis, and research agenda. *Journal of Business Research,* 124, 389-404. doi: 10.1016/J.JBUSRES.2020.10.044.

Paschen, J., Wilson, M., & Ferreira, J. J. (2020). Collaborative intelligence: How human and artificial intelligence create value along the B2B sales funnel. *Business Horizons,* 63(3), 403-414. doi: 10.1016/J.BUSHOR.2020.01.003.

Saifol Hosseini, F., Mohammadi nejad, A., Moghaddas, R. (2015). Comparing Forecasting Ability of Artificial Neural Networks and ARIMA Methods in Forecasting of Iran's Leather and Skin Exports, *Quarterly scientific-research journal of agricultural economics research,* 7(26), 125-143. ( in persian).

Salehi, M. (2016). Investigating the Impact of E-Commerce on the Development of Iranian Pistachio Exports. *International conference of new horizons in management and accounting sciences*, *economics and entrepreneurship,* 1175. ( in persian), doi: https://civilica.com/doc/567117.

Shahabadi, A., Amiri, B. & Sari gol, S. (2016). Identify Factors Affecting the Exports and Imports of NAM Member Countries with Emphasis on Institutional Governance Index. *Biannual Peer Review Journal of Business Strategies*. 22(2), 57-72. (in persian). doi: magiran.com/p1574361.

Sohrabpour, V., Oghazi, P., Toorajipour, R. & Nazarpour, A. (2021). Export sales forecasting using artificial intelligence. *Technological Forecasting and Social Change*, 163, 120480. doi: 10.1016/J.TECHFORE.2020.120480.

Syam, N. & Sharma, A. (2018). Waiting for a sales renaissance in the fourth industrial revolution: Machine learning and artificial intelligence in sales research and practice. *Industrial marketing management*, 69, 135-146. doi: 10. 1016/ J. INDMARMAN. 2017. 12. 019.

Tafaghodi, H.R., Ramazanian, M.R., Yakideh, K., & Akbari, M. (2020). Exploring the Role of the Ecosystem of Industrial Goods Export in Private Sector Businesses. Journal of Business Management, 12(2), 315-334. (in Persian) doi: https://civilica.com/doc/1386933/.

Tussyadiah, I. (2020). A review of research into automation in tourism: Launching the Annals of Tourism Research Curated Collection on Artificial Intelligence and Robotics in Tourism. *Annals of Tourism Research*, 81, 102883. doi: 10.1016/J.ANNALS.2020.102883.

Urrutia, J. D., Abdul, A. M., & Atienza, J. B. E. (2019, December). Forecasting Philippines imports and exports using Bayesian artificial neural network and autoregressive integrated moving average. *In AIP Conference Proceedings* (Vol. 2192, No. 1, p. 090015). AIP Publishing LLC.

Zare Mehrjerdi, M.R., Mehrabi Boshrabadi, H., Nezamabadi pour, H. & Tohidi, A. (2015). Evaluation of Artificial Neural Network-Panel Data Hybrid Model in Predicting Iran's Dried Fruits Export Prices. *Quarterly Journal of Quantitative Economics*, 12 (3), 95–116. (in persian), doi: magiran.com/p1518322.

Zarra nejad, M., Feghh Majid, A., Rezayi, R. (2008). Forecasting exchange rate using artificial neural network and ARIMA model, *Quantitative Economics Quarterly (former economic reviews),* 5(4), 107-130, (in persian).

Zhang, G., Patuwo, B. E., & Hu, M. Y. (1998). Forecasting with artificial neural networks: The state of the Art. International journal of forecasting, 14(1), 35-62.

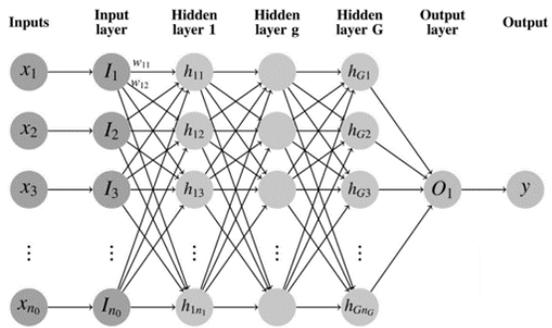

**Figure 1: Multi-Layer Perceptron Neural Network Structure**

Source **:**( Martínez-Comesaña et al., 2021)

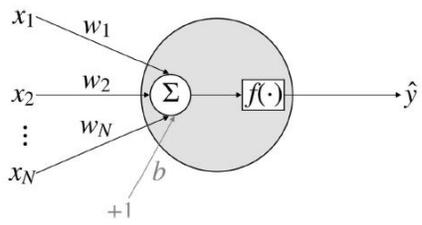

**Figure 2: Artificial Neuron Structure**

Source: ( Hoffmann, Bizarria & Bizarria, 2020)

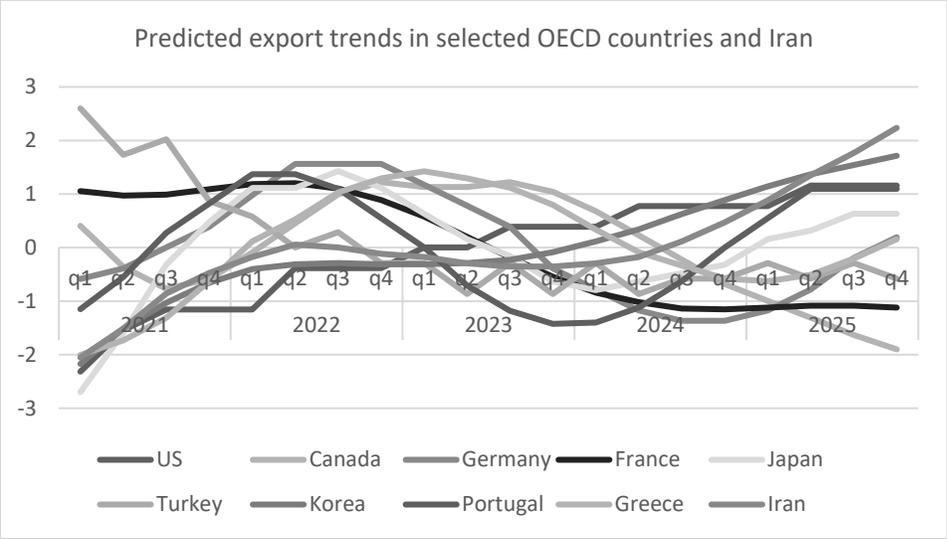

**Figure 3.Forecasted export trends in selected OECD countries and Iran**

Source: Author's Calculations

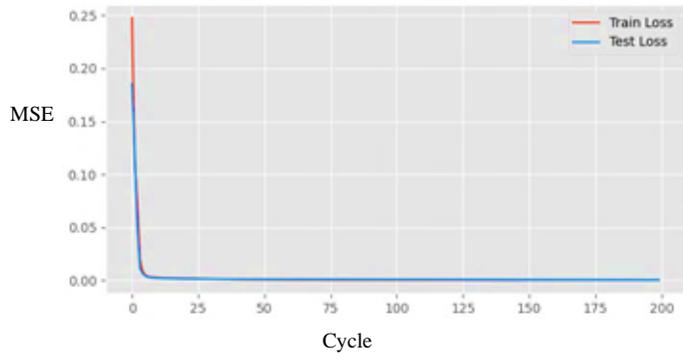

MSE

Cycle

**Figure 4. Change of MSE index with the number of training cycles.**

Source: Author's Calculations

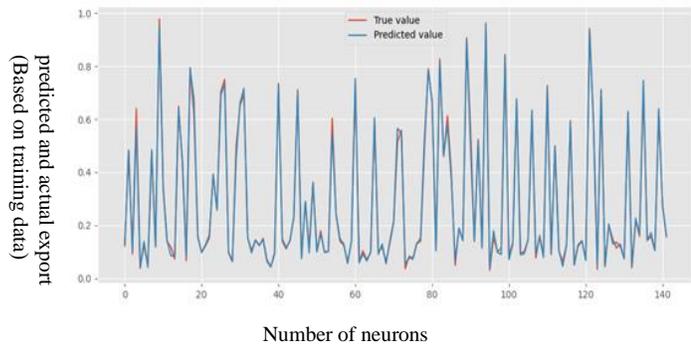

**Figure 5. Comparison of predicted vs actual export**

Source: Author's Calculations

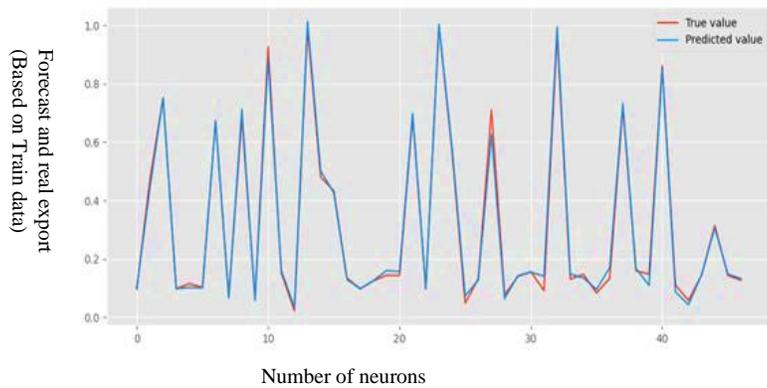

**Figure 6. Forecast and real export behavior**

Source: Author's Calculations

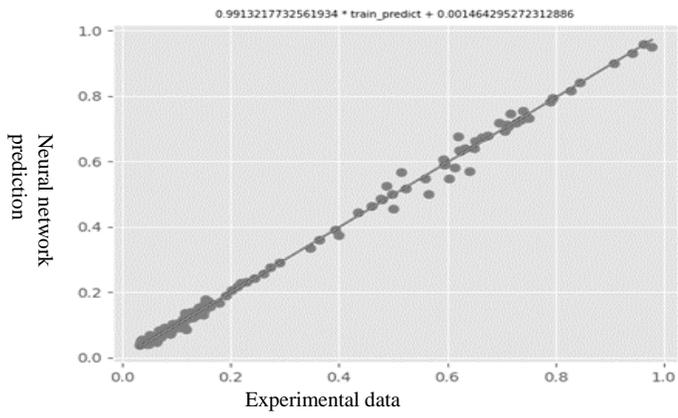

**Figure 7. Correlation between network predicted and actual viscosity values for training data.**

Source: Author's Calculations

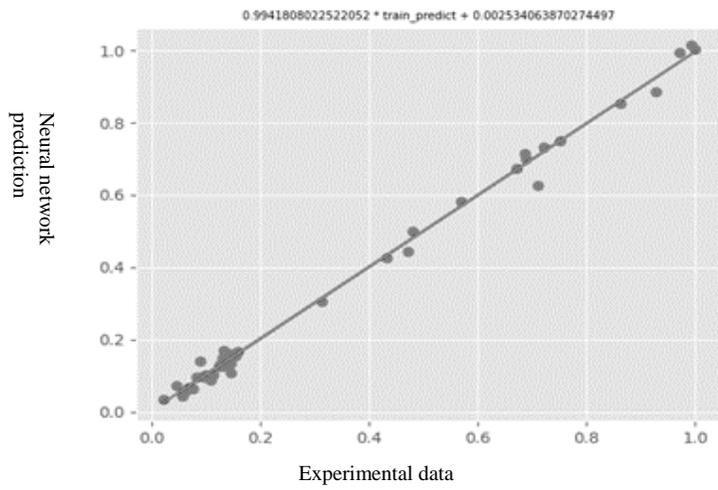

**Figure 8: Correlation between network predicted and actual viscosity values for test data**

Source: Author's Calculations

**Table 1. Forecasting error of MLP neural network test data**

|          | MSE         | MAE         |
|----------|-------------|-------------|
| US       | 0.000160188 | 0.008152633 |
| Canada   | 0.000730485 | 0.016155781 |
| Germany  | 0.000146367 | 0.008314841 |
| France   | 0.000140941 | 0.008552994 |
| Japan    | 0.000322108 | 0.010138276 |
| Turkey   | 0.000140084 | 0.006341856 |
| Korea    | 0.000413728 | 0.009945435 |
| Portugal | 0.000543046 | 0.014525768 |
| Greece   | 0.000500873 | 0.011423992 |
| Iran     | 0.000451706 | 0.014306198 |

Source: Author's Calculations

**Table 2. Forecasting error of MLP neural network Train data**

|  | MSE | MAE |
|---|---|---|
| US | 0.000207454 | 0.008166077 |
| Canada | 0.000562579 | 0.013291225 |
| Germany | 0.000186213 | 0.008206005 |
| France | 0.000194299 | 0.008706162 |
| Japan | 0.000409818 | 0.01144629 |
| Turkey | 0.000200517 | 0.007654217 |
| Korea | 0.000289028 | 0.009143234 |
| Portugal | 0.000268253 | 0.011002455 |
| Greece | 0.000242448 | 0.009860145 |
| Iran | 0.000255029 | 0.010063278 |

Source: Author's Calculations

**Table 3. Forecasting the amount of exports based on the multilayer perceptron (MLP) neural network (current US$)**

|  |  | US | Canada | Germany | France | Japan | Turkey | Korea | Portugal | Greece | Iran |
|---|---|---|---|---|---|---|---|---|---|---|---|
| 2021 | q1 | 2.43E+12 | 4.39E+11 | 1.69E+12 | 6.77E+11 | 8.64E+11 | 2.24E+11 | 7.06E+11 | 9.78E+10 | 7.39E+10 | 7.61E+10 |
|  | q2 | 2.45E+12 | 4.22E+11 | 1.7E+12 | 6.72E+11 | 8.71E+11 | 2.21E+11 | 7.29E+11 | 1E+11 | 7.58E+10 | 8.53E+10 |
|  | q3 | 2.46E+12 | 4.13E+11 | 1.72E+12 | 6.73E+11 | 8.79E+11 | 2.22E+11 | 7.46E+11 | 1.03E+11 | 7.87E+10 | 9.64E+10 |
|  | q4 | 2.46E+12 | 4.17E+11 | 1.74E+12 | 6.79E+11 | 8.84E+11 | 2.18E+11 | 7.59E+11 | 1.05E+11 | 8.34E+10 | 1.03E+11 |
| 2022 | q1 | 2.46E+12 | 4.28E+11 | 1.77E+12 | 6.85E+11 | 8.88E+11 | 2.17E+11 | 7.68E+11 | 1.07E+11 | 8.85E+10 | 1.08E+11 |
|  | q2 | 2.48E+12 | 4.4E+11 | 1.8E+12 | 6.86E+11 | 8.88E+11 | 2.15E+11 | 7.71E+11 | 1.07E+11 | 9.13E+10 | 1.12E+11 |
|  | q3 | 2.48E+12 | 4.52E+11 | 1.8E+12 | 6.8E+11 | 8.9E+11 | 2.16E+11 | 7.72E+11 | 1.06E+11 | 9.48E+10 | 1.11E+11 |
|  | q4 | 2.48E+12 | 4.57E+11 | 1.8E+12 | 6.67E+11 | 8.88E+11 | 2.14E+11 | 7.71E+11 | 1.04E+11 | 9.66E+10 | 1.09E+11 |
| 2023 | q1 | 2.49E+12 | 4.55E+11 | 1.78E+12 | 6.48E+11 | 8.85E+11 | 2.14E+11 | 7.71E+11 | 1.02E+11 | 9.75E+10 | 1.08E+11 |
|  | q2 | 2.49E+12 | 4.55E+11 | 1.76E+12 | 6.25E+11 | 8.82E+11 | 2.12E+11 | 7.72E+11 | 9.94E+10 | 9.66E+10 | 1.06E+11 |
|  | q3 | 2.5E+12 | 4.57E+11 | 1.74E+12 | 6.02E+11 | 8.8E+11 | 2.14E+11 | 7.74E+11 | 9.77E+10 | 9.54E+10 | 1.05E+11 |
|  | q4 | 2.5E+12 | 4.53E+11 | 1.7E+12 | 5.8E+11 | 8.77E+11 | 2.12E+11 | 7.79E+11 | 9.68E+10 | 9.32E+10 | 1.05E+11 |
| 2024 | q1 | 2.5E+12 | 4.45E+11 | 1.68E+12 | 5.62E+11 | 8.76E+11 | 2.14E+11 | 7.86E+11 | 9.69E+10 | 9E+10 | 1.06E+11 |
|  | q2 | 2.51E+12 | 4.35E+11 | 1.66E+12 | 5.51E+11 | 8.77E+11 | 2.12E+11 | 7.94E+11 | 9.79E+10 | 8.71E+10 | 1.08E+11 |
|  | q3 | 2.51E+12 | 4.25E+11 | 1.65E+12 | 5.44E+11 | 8.78E+11 | 2.13E+11 | 8.04E+11 | 9.97E+10 | 8.54E+10 | 1.13E+11 |
|  | q4 | 2.51E+12 | 4.15E+11 | 1.65E+12 | 5.43E+11 | 8.79E+11 | 2.13E+11 | 8.13E+11 | 1.02E+11 | 8.37E+10 | 1.19E+11 |
| 2025 | q1 | 2.51E+12 | 4.08E+11 | 1.66E+12 | 5.45E+11 | 8.82E+11 | 2.14E+11 | 8.22E+11 | 1.04E+11 | 8.34E+10 | 1.26E+11 |
|  | q2 | 2.52E+12 | 4.01E+11 | 1.68E+12 | 5.47E+11 | 8.83E+11 | 2.13E+11 | 8.3E+11 | 1.06E+11 | 8.42E+10 | 1.34E+11 |
|  | q3 | 2.52E+12 | 3.94E+11 | 1.71E+12 | 5.47E+11 | 8.85E+11 | 2.14E+11 | 8.36E+11 | 1.06E+11 | 8.63E+10 | 1.41E+11 |
|  | q4 | 2.52E+12 | 3.88E+11 | 1.73E+12 | 5.45E+11 | 8.85E+11 | 2.13E+11 | 8.42E+11 | 1.06E+11 | 8.88E+10 | 1.49E+11 |

Source: Author's Calculations